\documentclass[fleqn,usenatbib]{mnras}

\usepackage{newtxtext,newtxmath}
\usepackage[T1]{fontenc}

\DeclareRobustCommand{\VAN}[3]{#2}
\let\VANthebibliography\thebibliography
\def\thebibliography{\DeclareRobustCommand{\VAN}[3]{##3}\VANthebibliography}

\usepackage{graphicx}	% Including figure files
\usepackage{amsmath}	% Advanced maths commands

\defcitealias{Tominaga2014}{T14}
\defcitealias{Abate2015b}{A15b}
\defcitealias{Mura2020}{Paper I}

\usepackage{booktabs}
\usepackage{relsize}
\usepackage{xspace}
\usepackage{subfigure}
\usepackage{listings}
\usepackage{pdflscape}
\usepackage{threeparttable}
\usepackage{float}
\usepackage{multirow}
\usepackage{orcidlink}

\newcommand{\micro}{$\xi_{t}$~}
\newcommand{\velo}{km\ s$^{-1}$~}
\newcommand{\Teff}{$T_{\rm eff}$}  
\newcommand{\logg}{$\log(g)$}

\newcommand{\chem}[2]{ $^{#1}{\rm{#2}}$}
\newcommand{\Fe}{\rm{[Fe/H]}}
\newcommand{\FFe}{\rm{[F/Fe]}}
\newcommand{\Ba}{\rm{[Ba/Fe]}}
\newcommand{\C}{[\rm{C/Fe]}}
\newcommand{\Eu}{[\rm{Eu/Fe]}}

\newcommand{\kms}{$\mathrm{km\ s^{-1}}$}
\newcommand{\Msun}{M$_\odot$}

\newcommand{\MIi}{M$_{1,i}$}
\newcommand{\MIIi}{M$_{2,i}$}
\newcommand{\Mpmz}{M$_{\rm{pmz}}$}
\newcommand{\Macc}{M$_{\rm{acc}}$}
\newcommand{\OPi}{P$_{i}$}

\def\gtsim {>\kern-1.2em\lower1.1ex\hbox{$\sim$}~}   % Greater than sim
\def\ltsim {<\kern-1.2em\lower1.1ex\hbox{$\sim$}~}   % Less than sim

\usepackage{soulpos}
\usepackage{xcolor}

\ulposdef{\todel}{%
    \rlap{\textcolor{yellow}{\rule[-.75ex]{\ulwidth}{2.5ex}}}%
    \rule[.45ex]{\ulwidth}{.1ex}%
}

\title[Hints on the nature of the first stars using fluorine]{Fluorine abundances in CEMP stars at the lowest metallicity: Hints on the nature of the first stars}

\author[Mura-Guzm\'an et al.]{
Aldo Mura-Guzm\'an \orcidlink{0000-0003-1711-1981},$^{1,2,3,4}$\thanks{E-mail:\href{mailto:aldo.muraguzman@anu.edu.au}{aldo.muraguzman@anu.edu.au}}
David Yong \orcidlink{0000-0002-6502-1406},$^{3,4}$
Chiaki Kobayashi \orcidlink{0000-0002-4343-0487},$^{5}$
Nozomu Tominaga \orcidlink{0000-0001-8537-3153},$^{6,7,8}$ \newauthor
Madeleine McKenzie \orcidlink{0000-0002-1715-1257},$^{3,4,9}$
Ricardo Salinas \orcidlink{0000-0002-1206-1930},$^{10}$
Gregory Mace \orcidlink{0000-0001-7875-6391},$^{11}$
Hwihyun Kim \orcidlink{0000-0003-4770-688X},$^{12}$ \newauthor
Daniel B. Zucker \orcidlink{0000-0003-1124-8477}$^{1,2,4}$\\
$^{1}$Macquarie University Faculty of Science and Engineering, School of Mathematical and Physical Sciences, Macquarie University, Sydney, NSW 2109, Australia\\
$^{2}$Research School of Astronomy and Astrophysics, Australian National University, Mount Stromlo Observatory, Cotter Road, Weston Creek, ACT 2611, Australia\\
$^{3}$Astrophysics and Space Technologies Research Centre, Macquarie University, Sydney, NSW 2109, Australia.\\
$^{4}$ARC Centre of Excellence for All Sky Astrophysics in 3 Dimensions (ASTRO 3D), Australia\\
$^{5}$Centre for Astrophysics Research, University of Hertfordshire, College Lane, Hatfield, AL10 9AB, UK\\
$^{6}$National Astronomical Observatory of Japan, National Institutes of Natural Sciences 2-21-1 Osawa, Mitaka, Tokyo 181-8588, Japan\\
$^{7}$Astronomical Science Program, Graduate Institute for Advanced Studies, SOKENDAI 2-21-1 Osawa, Mitaka, Tokyo 181-8588, Japan\\
$^{8}$Department of Physics, Faculty of Science and Engineering, Konan University, 8-9-1 Okamoto, Kobe, Hyogo 658-8501, Japan\\
$^{9}$Carnegie Science Observatories, 813 Santa Barbara St., Pasadena, CA 91101, USA\\
$^{10}$Nicolaus Copernicus Astronomical Center, Polish Academy of Sciences, Bartycka 18, 00-716 Warszawa, Poland\\
$^{11}$Department of Astronomy and McDonald Observatory, University of Texas at Austin, 2515 Speedway, Stop C1400, Austin, Texas 78712-1205, USA\\
$^{11}$Gemini Observatory/NSF’s NOIRLab, 950 N Cherry Avenue, Tucson, AZ 85719, USA
}

\date{Accepted XXX. Received YYY; in original form ZZZ}

\pubyear{2024}

\begin{document}
\label{firstpage}
\pagerange{\pageref{firstpage}--\pageref{lastpage}}
\maketitle

\begin{abstract}
In the last decade, the available measurements of fluorine abundance have increased significantly, providing additional information on the chemical evolution of our Galaxy and details on complex stellar nucleosynthesis processes. However, the observational challenges to obtain stellar F abundances favor samples with higher metallicities, resulting in a scarcity of measurements at low-metallicity ($\Fe<-2.0$).
We present F abundances and upper limits in 7 carbon enhanced metal-poor (CEMP) stars observed with the Immersion Grating Infrared Spectrometer (IGRINS), at the Gemini-South telescope. These new observations delivered high-resolution, high signal-to-noise ratio, infrared spectra allowing us to probe significantly deeper into the metal-poor regime and the cosmic origin of F. This work presents the results of our observations, including two 2-sigma detections and five upper limits in a variety of CEMP stars. Arguably the most important result is for CS 29498-0043, a CEMP-no star at [Fe/H] = $-3.87$ with a F detection of [F/Fe] = $+2.0\pm 0.4$, the lowest metallicity star (more than a factor of 10 lower in metallicity than the next detection) with observed F abundance to date.
This measurement allowed us to differentiate between two zero metallicity Population III (Pop III) progenitors: one involving He-burning with primary N in Wolf–Rayet stars, and the other suggesting H-burning during hypernova explosions. Our measured value is in better agreement with the latter scenario.
This detection represents a pilot, and pioneering study demonstrating the power of F to explore the nature and properties of the first chemical enrichment from Pop III stars.
\end{abstract}


%%%%%%%%%%%%%%%%%%%%%%%%%%%%%%%%%%%%%%%%%%%%%
\section{Introduction}\label{Intro_CEMP2}
%%%%%%%%%%%%%%%%%%%%%%%%%%%%%%%%%%%%%%%%%%%%%

The first stars to have formed in the Universe contained only primordial material; hydrogen, helium, and trace amounts of lithium. For these Population III (hereafter Pop III) zero metallicity stars, the initial mass function was likely top-heavy and therefore dominated by massive stars \citep{Bromm2004,Frebel2015,Ishigaki2018,Sharda2022}.
However, the assumption of solar-scaled abundances (such as C and O) for a metal-poor initial mass function (IMF) may not be appropriate, and a different composition could change the profile of the IMF from top- to bottom-heavy \citep[see e.g.,][]{sharda23}.
In a top-heavy IMF scenario, this first generation of stars was massive and has long since died; the most massive ones as core-collapse supernova (SN). While we are unlikely to directly observe Pop III stars, we can still examine their masses, explosion energies, rotational velocities, and other physical properties affecting their nucleosynthetic yields in the Galactic archaeology.
In the absence of direct age measurements, which are challenging for individual stars, the chemical abundance profile is the best available proxy for a star's age. Using iron as a proxy for metallicity, the oldest most chemically primitive stars contain the nucleosynthetic signatures of the first generation of stars. Thus, the most metal-poor stars offer observational insights into the nature of the first generation of stars and the first chemical enrichment events in the early Universe. 

Among the low-metallicity, low-mass, long-lived stars still visible today, it has become clear that the fraction of metal-poor stars with a significant C enrichment ([C/Fe] $>0.7$), increases with decreasing metallicity \citep[e.g.,][]{Beers2005, Aoki2007,Placco2014,Hansen19}. Formed from gas clouds enriched with the remnants of previous stellar generations or chemically altered by accreted material from a companion \citep{Lucatello2005}, carbon enhanced metal-poor (CEMP) stars provide a unique window into the chemical landscape of the early Universe.

The information about the nucleosynthesis mechanisms that took place in the progenitors of CEMP stars, the first stars, can be accessed by studying their chemical composition. In addition to the C over-abundances in CEMP stars, enhancements in Ba and Eu  -- from the slow ($s$-) and rapid ($r$-) neutron capture processes, respectively -- are used to classify them into four different types: i) CEMP-$s$, ii) CEMP-$r$, iii) CEMP-$r/s$, and iv) CEMP-no.\footnote{CEMP-$s$: ([Ba/Fe]>0.5, [Ba/Eu]>0), CEMP-$r$: ([Eu/Fe]>1.0, [Ba/Eu]<0), CEMP-$r/s$: ([Ba/Fe]>0.5, [Ba/Eu]>0, [Eu/Fe]>1.0), and CEMP-no: ([Ba/Fe]<0.5, [Eu/Fe]<1.0). Classification criteria vary slightly among other authors, see e.g., \cite{Beers2005,Aoki2007,Masseron2010}} These distinct chemical patterns observed in their stellar atmospheres suggest various formation scenarios, including 1) faint SN with small ejection of Fe; 2) dark SN without Fe ejection; 3) mass loss from a rotating massive star; 4) mass transfer from an asymptotic giant branch (AGB) star companion; 5) self-enrichment; and 6) external enrichment \citep[see e.g.,][for a review on the topic]{Nomoto2013}.

Radial velocity monitoring of the CEMP-$s$ class has revealed that these are binaries \citep{Lucatello2005}. For these objects, the carbon and $s$-process element excess were obtained via mass transfer from an AGB companion. Recently, \cite{Arentsen2019} also suggested a high binary fraction in CEMP-$no$ stars, although the authors are careful to note that \textit{binarity does not equate to mass transfer}. Clearly, more efforts are needed to constrain the formation channels for CEMP stars.
Understanding the relative importance and the frequency of these processes remains a significant challenge in modern astrophysics. Therefore, resolving the different formation channels is key to understanding the complex history of early stellar evolution, and ultimately, the chemical enrichment of the early Galaxy.\\

Fluorine is rare and fragile, and arguably one of the most interesting elements. With only one stable isotope, $^{19}$F, fluorine is also one of the least abundant elements in the solar system (among the elements less massive than Zn). In the last decade, increasing interest to understand the cosmic origin of F has led to more observations \citep[e.g.,][]{Jonsson2014b,Guerco2019,Ryde2020,Mura2020} and refinements in theoretical modeling of nucleosynthesis and Galactic chemical evolution \cite[e.g.,][]{Lugaro2004,Karakas2010,Kobayashi2011a,Prantzos2018,Womack2023}. 
Very recently, \citet{Nandakumar2023} investigated the evolution of F at higher metallicities ($-0.9 < \Fe < 0.25$ dex) and recommended a set of vibrational-rotational HF lines to be used for abundance analysis. They note that the abundances derived from the strongest lines show significant trends with the stellar parameters. However, these issues are not observed in the metal-poor regime such as is the case of this work (i.e., $-3.9 < \Fe < -2.2$). In another recent study by \citet{Seshashayana20240a,Seshashayana20240b}, the authors obtained F abundances as well as Ce as a tracer for $s$-process from open clusters using the high-resolution near-infrared GIANO-B instrument at the Telescopio Nazionale Galileo. Comparing their results to Galactic chemical evolution models, their findings suggest that the cosmic origins of F need AGB and massive stars (including fast rotators) as main production source.

Being highly sensitive to the conditions where formed, F can serve as a tool for constraining nucleosynthesis and chemical evolution models. However, the complex nucleosynthesis channels for the production of $^{19}$F and the volatility of the isotope to be easily destroyed have challenged theoretical models to describe the observations. %
\citet{Franco2021} reported the high F abundance in NGP–190387, a lensed galaxy at $z=4.4$. Due to its redshift, the galaxy is observed as it was 1.3 billion years after the Big Bang. 
Therefore, the detection of F in this galaxy indicates that AGB stars are unlikely to be the source of F production given their typical lifetimes. Instead, the authors suggest that Wolf--Rayet stars are responsible for the observed F abundances in such a young galaxy.

Fluorine is primarily formed in three astrophysical sites: core-collapse SNe (CCSNe) \citep{Woosley1995}, Wolf--Rayet stars \citep{Meynet2000}, and AGB stars \citep{Goriely2000A, Busso1999, Lugaro2004,Renda2004,Cristallo2011}. 
%
%CCSN
The F production in CCSN is dominated by neutrino spallation \citep{Woosley1988} where neutrinos produced from the explosion interact with the Ne-rich layer of the dying star producing a significant amount of F. \citet{Kobayashi2011b} showed that this can explain the evolution of F abundance in the Milky Way. Note that, however, such large F enhancements from the neutrino spallation process are yet to be observationally confirmed. Additionally, prior to the explosion, F is produced through CNO burning in the He shell \citep{Heger2010}.
%
%AGB stars
In AGB stars with M < 4\Msun, F is produced in the He intershell  \citep[see e.g.,][]{Forestini1992,Jorissen1992,Mowlavi1998}, and it is a companion for C and $s$-process elements \citep{Mowlavi1998}. AGB stars are the only observationally confirmed site of F production \citep{Jorissen1992,Werner2005,Zhang2005,Pandey2006,Schuler2007,Abia2010,Lucatello2011}. In addition, AGB nucleosynthesis models exhibit strong dependence on the initial stellar mass \citep{Karakas2010}. Thus, expanding fluorine observations in the most metal-poor stars and comparing them with detailed model predictions will allow us to place tighter constraints on the initial mass function in the early Universe
%
%Massive stars
Massive stars produce F during Helium burning. Once formed, the F isotope needs to be removed from the interior before being destroyed. When massive stars enter the Wolf--Rayet phase, the external layers containing the processed material are ejected into the interstellar medium through strong stellar winds \citep{Meynet2000,Limongi2018}. Significant enhancements in F abundances are expected if rotation is included in the models.
Therefore it is very important to measure F abundances across cosmic time, or for a wide range of metallicity of stars in the Milky Way.

The diagnosis of F abundances in late-type (FGK) stars commonly relies on the only unblended molecular absorption of the HF (1-0) R 9 line at 23358.3\AA, in the near-infrared \citep[see e.g.,][]{Jorissen1992,Jonsson2014a}. The measurements of F abundance at low metallicities via HF have been restricted by detection limitations, such as the weakness of the line, temperature sensitivity, and the requirement of a high signal-to-noise ratio (SNR) and high-resolution spectra \citep{Jonsson2014a,Ryde2020,Mura2020}. In CEMP stars, four measurements of F abundances have been reported only in the CEMP-$s$ class: two detections and eight upper limits by \cite{Lucatello2011}; one detection by \cite{Schuler2007}; and recently, one detection and one upper limit by \citet[][hereafter \citetalias{Mura2020}]{Mura2020}.
Additionally, \cite{Li2013} obtained F abundances detections (2) and upper limits (5) in seven halo field metal-poor stars (not CEMP stars) from [Fe/H] = $-1.56$ to $-2.13$, using CRIRES\footnote{\url{https://www.eso.org/sci/facilities/paranal/instruments/crires.html}} \citep{CRIRES2004} at the Very Large Telescope. They also considered the effects of three-dimensional model atmospheres on the derived F abundances and found them to be insignificant for their program stars.

Although the current observational sample of F abundances in CEMP stars is limited, the derived abundances (and estimated upper limits) have revealed some discrepancies with theoretical models. In \citetalias{Mura2020} we compared our observations, along with those from \cite{Lucatello2011}, against theoretical predictions of AGB nucleosynthesis and binary stellar evolution from \cite{Abate2015b} -- which are based on the work from \citet{Lugaro2012}. From our results in \citetalias{Mura2020}, we found that while some F abundances in CEMP-$s$ stars are in good agreement with the theoretical predictions from \cite{Abate2015b}, the remaining portion of the sample (only upper limits), predominantly consisting of CEMP-$r/s$ stars, show lower F content than the model predictions. These results reflect the discussion from \cite{Abate2015b} which indicates that their models tend to overproduce light elements while attempting to reproduce the observed abundances of $s$- and $r$-process elements simultaneously. The inability of the model to reproduce the chemical pattern observed in CEMP-$r/s$ stars may suggest alternative formation scenarios and nucleosynthesis processes. Therefore, it is extremely important to measure F abundances in CEMP-no stars, which are unlikely to be affected by enrichment from AGB companion stars.

We present new observations for 7 CEMP stars using spectra obtained from the Immersion Grating Infrared Spectrometer \citep[IGRINS;][]{IGRINS1,IGRINS2,Mace2018} at the Gemini-S telescope. The target selection and overview, and observations and data reduction are described in Sections \ref{Target_Selection_Overview}, and \ref{Observations_CEMP2}, respectively. Section \ref{Analisys_CEMP2}, presents the analysis of the data, followed by the obtained results in Section \ref{Results_CEMP2}. A discussion is open in Sect. \ref{Discussion_CEMP2} finalizing with the conclusions of this work in Sect. \ref{Conclusions_CEMP2}.

%%%%%%%%%%%%%%%%%%%%%%%%%%%%%%%%%%%%%%%%%%%%%
\section{Target Selection and Overview}\label{Target_Selection_Overview}
%%%%%%%%%%%%%%%%%%%%%%%%%%%%%%%%%%%%%%%%%%%%%

The selection of the targets was made following the same criteria used in \citetalias{Mura2020}.
Given the very low metallicity of the stars under study, a high F abundance is necessary to produce a strong enough HF absorption to be detected. In addition to a high F content, stellar parameters have to be restricted as well, due to the sensitivity of the HF line to \Teff\, and \logg,  in order to increase the chances of detection (see more details in Section \ref{detection_limit}).

HE 1305+0007 and HE 1523-1155 were selected considering the predicted F abundances from the theoretical models of AGB nucleosynthesis and binary evolution from \cite{Abate2015b}, as in \citetalias{Mura2020}.
CS 29498-0043 was selected due to the significant F over-abundance predicted by zero metallicity Pop III  SN nucleosynthesis models from \cite{Tominaga2014}. The observations of F abundances in these objects, although challenging due to the weakness of the line, provide an excellent test and important constraints for state-of-the-art nucleosynthesis models in faint SNe and AGBs at very low metallicity. The remaining targets, CS 29502-0092, HD 126587, HE 0414-0343, and HE 1116-0634, have not been analyzed with nucleosynthesis models so far, and therefore, there are no custom-made predictions of F abundances in the literature for these stars. However, these objects are very metal-poor with stellar parameters (i.e., low \Teff) such that F measurements may be possible. As these objects were observable at Gemini-S, they presented a good opportunity for observations and to expand the current known sample in the low-metallicity regime.\\

Our selected targets include 7 CEMP stars encompassing CEMP-$s$, -$r/s$, and -$no$ types. A brief description of the sample using compiled information from \cite{Yoon2016a} and references therein (otherwise specified), as well as theoretical models from \cite{Abate2015b} and \cite{Tominaga2014} (when available), is given as follows:

\begin{itemize}
    \item[$\star$]\textit{\textbf{HE 1523-1155:}} CEMP-$s$ star in a binary system with a metallicity of $\Fe= -2.20$ and enhanced C and Ba abundances of $\C=+1.94$ and $\Ba=+1.8$, respectively. No Eu abundance has been reported for this object. The physical parameters of the model from \cite{Abate2015b} that best fit the chemical profile of HE 1523-1155 are: (i) initial masses \MIi\ $=1.7$ \Msun\ and \MIIi\ $=0.76$ \Msun; (ii) initial orbital period \OPi\ $= 1.75\times 10^5$ days; (iii) mass of the partial mixing zone \Mpmz\ $=4 \times 10^{-3}$ \Msun; and accreted mass (iv) \Macc\ $=0.11$ \Msun. The predicted F abundance from that model is \FFe$_{\rm{Abate}}=2.24$.

    \item[$\star$]\textit{\textbf{HE 1305+0007:}} CEMP-$r/s$ star with mean metallicity $\Fe=-2.28$ \citep[see details in][and references therein]{Abate2015b}, over-abundances in both, $r$- and $s$-process elements $\Ba=+2.32$ and $\Eu=+1.97$, respectively, and $\C=+1.90$. The models by \cite{Abate2015b} predict a F abundance of \FFe$_{\rm{Abate}}=+2.27$. The model parameters are: (i) \MIi\ $=1.5$ \Msun and \MIIi\ $=0.54$ \Msun; (ii) \OPi\ $= 2.42\times 10^4$ days; (iii) \Mpmz $=6.66 \times 10^{-3}$ \Msun; (iv) \Macc$=0.32$ \Msun. HE 1305+0007 is a binary candidate, although inconsistencies in the radial velocities measurements from multiple sources cannot provide an orbital solution. However, as mentioned in \citetalias{Mura2020}, the model does not agree with the observed F abundances, nor provides a good fit of the $s$- and $r$-process elements simultaneously.

    \item[$\star$]\textit{\textbf{HE 0414-0343:}} CEMP-$r/s$ star in a binary system. With a metallicity of $\Fe=-2.24$. The reported relative abundance ratios for C, Ba, and Eu from literature are $\C= +1.60$, $\Ba= +1.87$, and $\Eu=+1.23$, respectively. HE 0414-0343 is not included in the  \cite{Abate2015b} modeling sample. However, \cite{Hollek2015} presents models that can reproduce the chemical distribution in HE 0414-0343 from a $>1.3$ \Msun\ AGB star in conjunction with a late mass transfer. While no explicit value for F abundance is shown in \cite{Hollek2015}, an inspection of their fig. 14.3 indicate a predicted fluorine content of A(F) $\sim +2.0$.

    \item[$\star$]\textit{\textbf{CS 29498-0043:}} Single CEMP-$no$ star with an extremely low-metallicity of $\Fe=-3.87$ \citep{Roederer2014b}, the lowest in our sample. The C abundance in this star is $\C=+3.06$. No enhancements in neutron capture elements are reported in the literature, $\Ba=-0.49$, $\Eu<+0.23$. CS 29498-0043 is of particular interest in this study due to the nature of the nucleosynthesis channel responsible for its chemical abundance distribution: A single faint SN progenitor of a Pop III star. \cite{Tominaga2014} presented two models (A and B in their work) that best fit the abundances observed in CS 29498-0043. Model A well reproduces the observed chemical composition using: (i) an explosion energy of E$_{51} = \rm{E}/(10^{51} \rm{erg}) =20$, (ii) remnant mass M$_{\rm rem}=5.27$ \Msun, (iii) an ejected mass of iron M$_{\rm ej}({\rm Fe})=9.10 \times 10^{-4}$ \Msun, and (iv) a ``low-density'' factor $f_\rho=1/2$ to enhance the entropy by mimicking aspherical explosions. Model B is also able to describe the chemical abundance distribution in CS 29498-0043 with similar physical parameters: (i)  E$_{51} = 20$, (ii) M$_{\rm rem}=5.11$ \Msun, (iii) M$_{\rm ej}({\rm Fe})=9.08 \times 10^{-4}$ \Msun, and (iv) $f_\rho=1/2$. Both predictions adopt the mixing-and-fallback models to imitate aspherical explosion \citep{Umeda2002}. The main difference between these two models is the mixing efficiency during hydrostatic burning of the progenitor star. 
    The progenitor star of Model A has no enhanced mixing while for Model B mixing is enhanced; both progenitor models are taken from \citet{iwamoto2005}.
    In Model A, the high N abundance of CS 29498-0043 may be explained by the self-enrichment due to the CNO cycle in the star, while in Model B, the N abundance is enhanced in the pre-supernova star. As a result, these models predict very different F abundances: $\FFe=-0.25$ and $\FFe=+3.02$ for models A and B, respectively. For the predicted $\FFe= +3.02$ value, synthetic spectra indicated that the HF should be clearly detectable, and thus this object was our highest priority target.

    \item[$\star$] The remaining stars in our sample are CEMP-$no$ stars which have not been included in either theoretical work from \cite{Abate2015b} nor \cite{Tominaga2014}. The stellar parameters in these stars suggested they may had the conditions for an observable HF line. Therefore, they were selected as opportunity targets given they were also visible during the observation period. CS 29502-0092, HD 126587, and HE 1116-0634 have metallicities of $\Fe= -2.28, -3.29, -3.73$, respectively. CS 29502-0092 is a single star with a C abundance of $\C=+1.46$. The Ba and Eu contents are very low, with $\Ba=-1.36$ and $\Eu<+0.11$. HD 126587, has similar metallicity to CS 29502-0092, although lower C, $\C=+0.85$, and higher Ba and Eu, $\Ba=-0.25$, $\Eu=+0.24$. The binary status is unknown for this star. HE 1116-0634 is the second lowest metallicity star in our sample (\Fe$=-3.73$), with a C abundance of \C$=+0.81$. The Ba abundance in this star is the lowest for the entire sample and no Eu content has been reported. The binary status of HE 1116-0634 has not been determined either.

\end{itemize}

%%%%%%%%%%%%%%%%%%%%%%%%%%%%%%%%%%%%%%%%%%%%%
\section{Observations and Data Reduction}\label{Observations_CEMP2}
%%%%%%%%%%%%%%%%%%%%%%%%%%%%%%%%%%%%%%%%%%%%%

The observations were made using IGRINS at the Gemini-S telescope over several years. Our scheduled observations for program GS-2020A-Q-212 were incomplete due to disruptions during the Covid-19 pandemic and later re-scheduled under program GS-2021A-Q-228. The exposure sequence for the observations follows an `ABBA' pattern along the 0.63" slit of the instrument. The total exposure time is composed of the number of individual exposure times per frame, providing high SNR of over 120 per resolution element in the $K$-band. The spectral resolution for all observations is $\sim$ 45\,000. The observation of each science target was followed by an A0V calibration star to later subtract telluric contamination from the earth's atmosphere \citep[see sections 3.4.1 and 3.4.2 from][]{sim14}. Further information on the observations is presented in Table \ref{t:obs_CEMP2}.

The final spectra used in the analysis were processed using the data reduction pipeline PLP2 \citep[plp, v 2.1,][]{lee17} as described in sect. 3 of \citetalias{Mura2020}.
For some of the targets, PLP2 failed to properly perform the telluric correction due to the lack of absorption features at these extremely-low metallicities. In those cases, we used the \textsc{iraf} \citep{IRAF1986,IRAF1993} task \verb+telluric+ to manually subtract the telluric contamination on the HF region at 2.33 $\mu$m. The normalization procedure was conducted using \textsc{iraf} task \verb+continuum+.

%\begin{landscape}
\begin{table*}
\begin{center}
    \begin{threeparttable}
        
\caption{Observing log from infrared spectroscopy by IGRINS.}\label{t:obs_CEMP2}
\begin{tabular}{lllrrccrc}
\hline
Star ID &       RA  &     Dec. &  $H$      &  $K$       &  Program ID &  Obs. Date &  Exp. Time &    $K$ SNR$^{\rm{b}}$  \\
        &    (2000) &    (2000)&  mag.$^{\rm{a}}$ &  mag.$^{\rm{a}}$  &	          &  (UT)      &  (s)       &                 \\
\hline
CS 29498-0043 & 21:03:51.85  &  -29:42:49.71 &  11.1    &  10.9    &  GS-2022B-DD-102 &  2022-10-18 &  957 $\times$ 8 & 220 \\
              &              &               &          &          &  GS-2021A-Q-228  &  2021-04-29 &  600 $\times$ 6 & 300 \\
HE 1305+0007  &  13:08:03.70 &  -00:08:44.11 &   9.8    &   9.6    &  GS-2020A-Q-212  &  2020-02-08 &  170 $\times$ 4 & 122 \\
              &              &               &          &          &  ---             &  2019-04-14 &  300 $\times$ 4 & 160 \\
HE 1523-1155  &  15:26:40.91 &  -12:05:45.54 &  10.9    &  10.8    &  GS-2021A-Q-228  &  2021-05-17 &  630 $\times$ 4 & 300 \\
CS 29502-0092 &  22:22:35.85 &  -01:38:23.91 &   9.7    &   9.6    &  GS-2020B-Q-315  &  2020-11-07 &  500 $\times$ 4 & 250 \\
HD 126587     &  14:27:00.31 &  -22:14:34.90 &   6.8    &   6.7    &  GS-2020B-Q-315  &  2021-01-05 &  8   $\times$ 8 & 247 \\
HE 0414-0343  &  04:17:16.40 &  -03:36:26.38 &   8.7    &   8.5    &  GS-2020B-Q-315  &  2020-11-15 &  250 $\times$ 4 & 650 \\
HE 1116-0634  &  11:18:35.81 &  -06:50:40.13 &   9.1    &   9.0    &  GS-2021A-Q-228  &  2021-01-19 &  120 $\times$ 4 & 136 \\
\hline
\end{tabular}
    \begin{tablenotes}
    \item[a] K magnitudes from 2MASS \citep{Cutri2003}.
    \item[b] $K-$band signal-to-noise ratios are indicated following the relation from IGRINS simple Exposure Time Calculator for Gemini (available at: \url{https://igrins-jj.firebaseapp.com/etc/simple}).
    \end{tablenotes}
        \end{threeparttable}
    \end{center}
\end{table*}

%\end{landscape}

%%%%%%%%%%%%%%%%%%%%%%%%%%%%%%%%%%%%%%%%%%%%%
\section{Data Analysis}\label{Analisys_CEMP2}
%%%%%%%%%%%%%%%%%%%%%%%%%%%%%%%%%%%%%%%%%%%%%
The determination of \Teff\ and \logg\ via spectroscopic analysis (e.g., excitation and ionization balance) is simply not possible in the IR for these objects due to the low number of Fe I absorption lines through the $H$- and $K$-band and the absence of Fe II lines. Moreover, this approach is also problematic due to non-local thermodynamic equilibrium (NLTE) effects \citep{lind12}. These issues complicate the determination of \Teff\ and \logg\ via traditional spectroscopic methods. An alternative option is using line depth ratios (LDR) and their relation with \Teff\ \citep{fukue15,afsar23}. Unfortunately, the low-metallicity nature of the sample under study in this work is incompatible with the metallicity regime of the IR LDR-\Teff\ methods available. Therefore, the stellar parameters for the sample were adopted from optical analysis available in the literature (see Table \ref{t:stellar_param_cemp2}).

\begin{table}
    \begin{center}
        \begin{threeparttable}
\caption{Stellar parameters, CEMP type, Binary status and references for the observed sample.}\label{t:stellar_param_cemp2}
\begin{tabular}{l
c@{\hspace{2mm}}
c@{\hspace{2mm}}
c@{\hspace{2mm}}
c@{\hspace{2mm}}
c@{\hspace{2mm}}
c@{\hspace{2mm}}
c@{\hspace{2mm}}}

\toprule
Star ID & \Teff & \logg & \micro       & \Fe&  Type &  Bin.$^{\rm{a}}$  & ref.$^{\rm{b}}$	\\
        & (K)   &       & (km s$^{-1}$)&    &       &            & \\
\midrule
CS 29498-0043 &  4440 &  0.50 &  1.54 &  $-3.87$ & no   & 1 &  1 \\
HE 1305+0007  &  4655 &  1.50 &  1.41 &  $-2.28$ & r/s  & - &  2,3 \\
CS 29502-0092 &  4820 &  1.50 &  1.30 &  $-3.30$ & no   & 1 &  1 \\
HD 126587 &      4640 &  1.00 &  1.40 &  $-3.29$ & no   & - &  1 \\
HE 0414-0343  &  4863 &  1.25 &  1.28 &  $-2.24$ & r/s  & 2 &  4 \\
HE 1116-0634  &  4400 &  0.10 &  2.40 &  $-3.73$ & no   & - &  5 \\
HE 1523-1155  &  4800 &  1.60 &  1.32 &  $-2.20$ & s    & 2 &  6 \\
\hline
\end{tabular}

        \begin{tablenotes}
        \item[a] Binary status: 1= single; 2=binary; from \cite{HansenT2016} (CEMP-no) and \cite{HansenC2016} (CEMP-$s$/-$rs$).
        \item[b] References: 1= \cite{Roederer2014b}; 2= \cite{Goswami2006}; 3= \cite{Beers2007}; 4= \cite{Hollek2015}; 5= \cite{Hollek2011}; 6= \cite{Aoki2007}.

        \end{tablenotes}
        
        \end{threeparttable}
    \end{center}

\end{table}

The abundance measurements were performed as indicated in \citetalias{Mura2020}, via spectrum-synthesis fitting using the local thermodynamic equilibrium (LTE) stellar line analysis program MOOG\footnote{\url{https://www.as.utexas.edu/~chris/moog.html}} \citep{sneden73} coupled with one-dimensional LTE model atmospheres from the \citet{Castelli2003} grid. Fluorine abundances were measured from the vibration–rotation transition of the HF molecular absorption line, HF (1-0) R 9, at 23358.329 \AA. The excitation potential and oscillator strength values used for HF (1-0) R 9, are  $\chi$ = 0.227 and $\log gf$\ $= -3.962$, respectively  \citep[and references therein]{Jonsson2014a}. Additionally, molecular features expected to have non-negligible contributions in the vicinity of the HF region were included in the line list, such as CO, CN, and their isotopologues \citep{Goorvitch1994,Sneden2014}. In order to account for a partial blending with $^{12}$C$^{17}$O line\footnote{IGRINS resolution R $\sim$ 45\,000 is sufficient to resolve the partial blending of $^{12}$C$^{17}$O with HF.} on the blue-end of used HF line, we adopted C abundances values from literature compilation by \citet[][and references therein]{Yoon2016a} for each star and adjust the abundances to match the observed spectra. The F abundances in the synthetic spectra were obtained using a chi-square fitting procedure, iteratively adjusting the F content in the synthetic spectra to fit the observed spectra as shown in Table \ref{t:Fabundances}. Since stellar parameters were adopted from literature compilations from \citet{Abate2015b} and \citet{Tominaga2014}, we considered the errors of our F measurements due to uncertainties in the stellar parameters assuming $\Delta$\Teff $=+100$ K, $\Delta$\logg $=+0.50$dex (cgs), $\Delta$\Fe $=+0.3$ dex, $\Delta$\micro $=+0.3$ \kms as typical individual uncertainty values added in quadrature.\\

\begin{figure}
\begin{center}
    		\includegraphics[width=\linewidth]{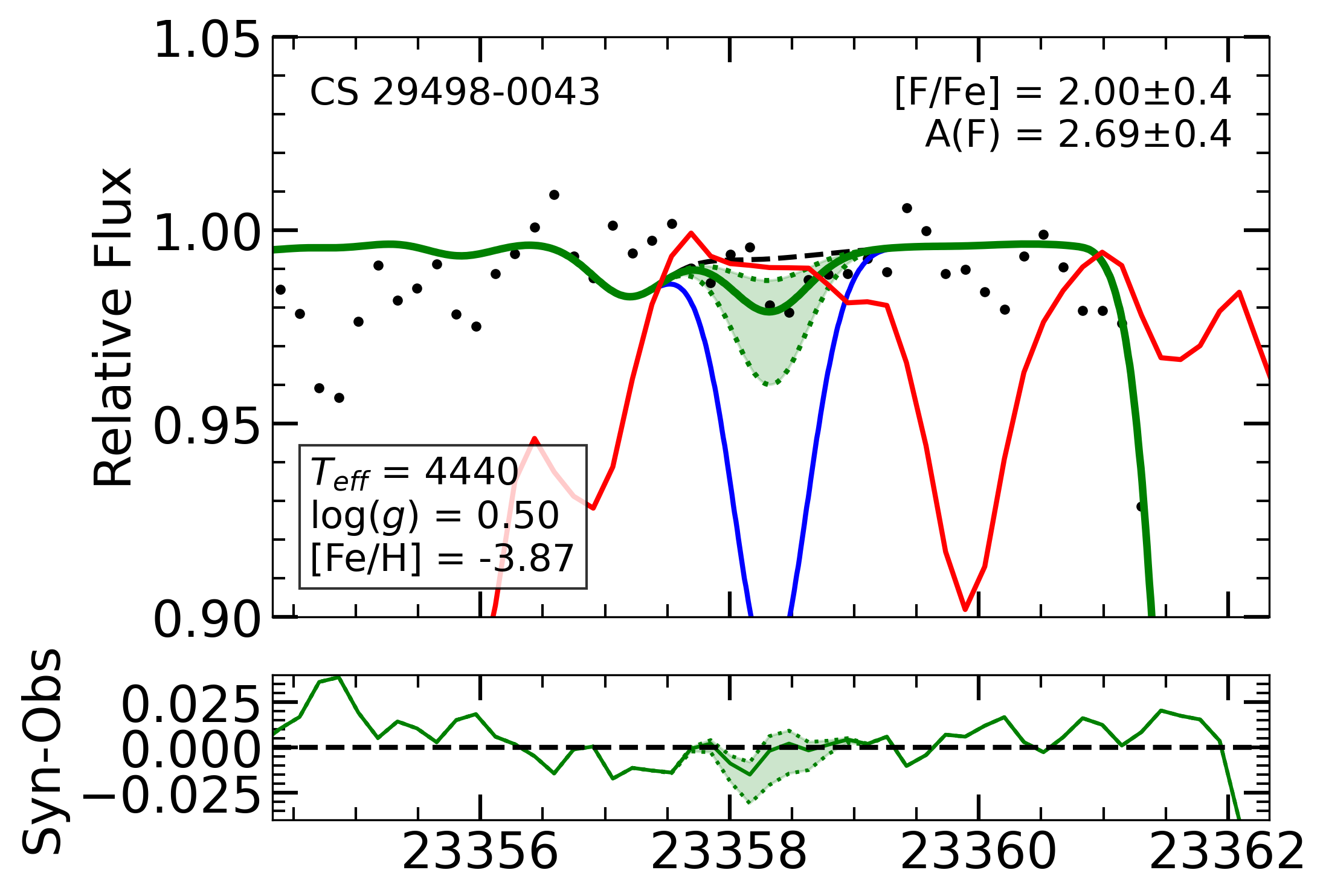}
          	\includegraphics[width=\linewidth]{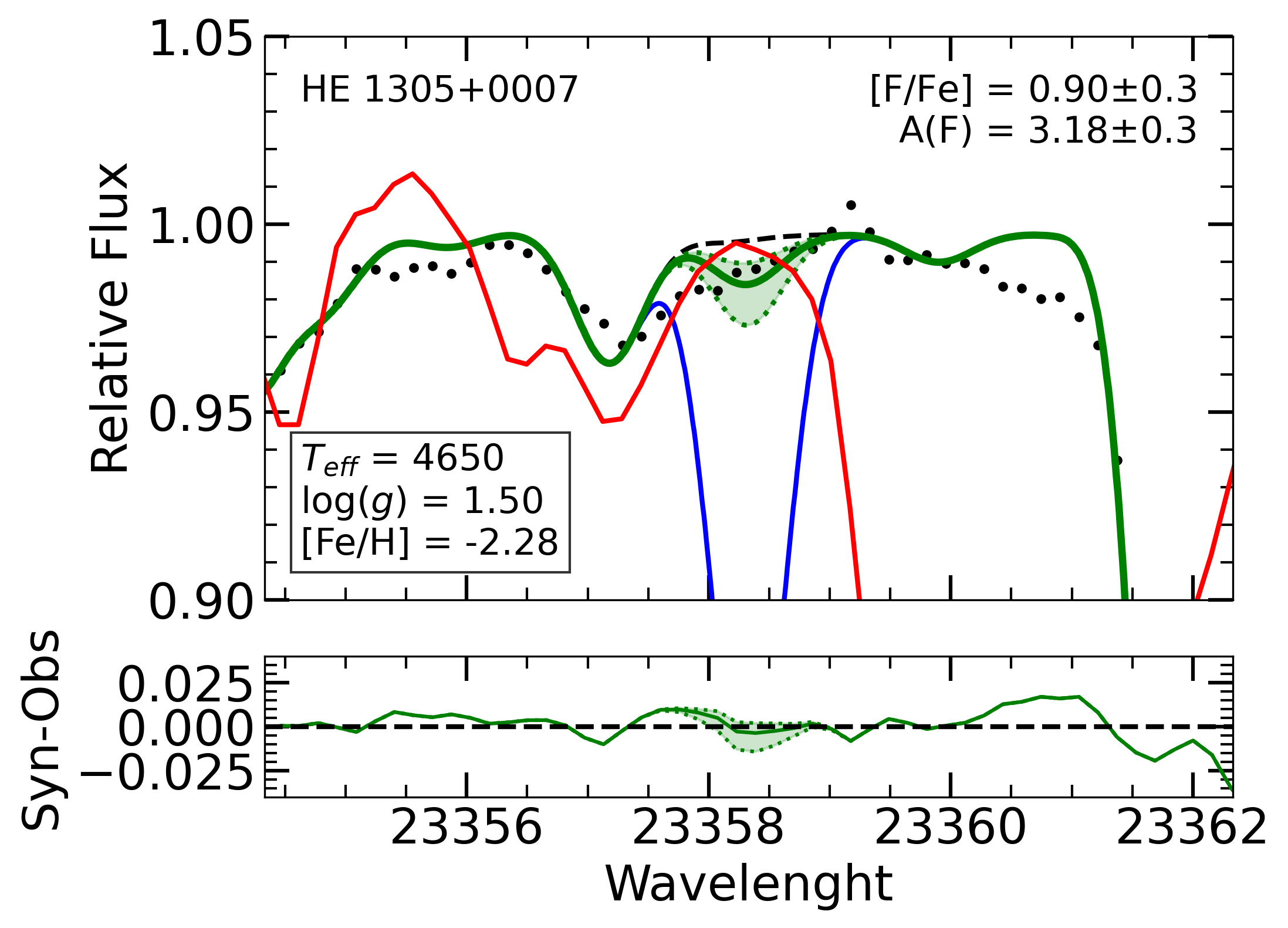}
			\caption[Synthetic spectra fitted to observed spectra from CS 29498-0043 (upper panel) and HE 1305+0007 (lower panel) in the HF region at $\sim 2.3\mu$m.]{Synthetic spectra fitted to observed spectra from CS 29498-0043 (upper panel) and HE 1305+0007 (lower panel) in the HF region at $\sim 2.3\mu$m. Observed spectra are presented in black points. Solid and dotted green lines describe the best-fit synthetic spectra and the uncertainty region (the green-shaded area) indicated in the top-right corner, respectively. The dashed black and solid blue lines represent no F absorption and the predicted absorption from theoretical models from \citet{Tominaga2014} and \citet{Abate2015b} for CS-29498-0043  (upper panel) and H 1305+0007 (lower panel), respectively. The synthetic spectra were produced using the stellar parameters shown in the bottom-left box. Additionally, uncorrected observed spectra are presented with solid red lines. Fitting residuals are displayed at the bottom of each synthesis.} \label{fig:detection}
\end{center}
\end{figure}

The HF absorption at 2.3$\mu$m becomes very weak at low metallicities ($\Fe <-2.0$), thus high-quality spectra (both spectral resolution and SNR) are mandatory to enable possible detections. However, despite a high SNR, an additional source of noise can be introduced via the telluric correction during the data reduction process. While the magnitude of the noise due to telluric correction is likely of the order of a few percent, such a contribution can be significant and challenging when the expected depths of the HF absorption line are of the same order. To avoid the telluric residuals on the HF line region, two stars in our sample with the potential of detection were re-observed, CS 29498-0043 and HE 1305+0007. These additional observations considered Earth's motion velocity shifts to displace the HF line into telluric-free regions. The high SNR and the absence of telluric contamination in the HF vicinity enabled the observation of 2-sigma F abundance detections in CS 29498-0043 and HE 1305+0007. The fitting of F abundances in these two stars are displayed in Fig. \ref{fig:detection}. In that figure, the telluric-corrected and telluric-contaminated spectra are shown in dotted black lines and solid red lines, respectively (both, corrected and uncorrected spectra are shifted to rest wavelengths in Fig. \ref{fig:detection}). The fitted synthetic spectra are generated using the indicated stellar parameters and F abundances presented in solid green lines. The effect on the synthesized spectra due to the calculated errors are highlighted in the region within dotted greern lines. A synthetic spectrum with no HF absorption (i.e., no F) and using the predicted theoretical F value are represented in a dashed black line and solid blue line, respectively. The velocity shift of the HF lines clearly isolates them from telluric contamination in CS 29498-0043 and HE 1305+0007. In addition, the residuals from the fitting are presented below each panel.

%#######################
%Fig "detections"
%#######################

%#######################
% Fig continuum regions 
%#######################
\begin{figure*}
\begin{center}
    		\includegraphics[width=.9\linewidth]{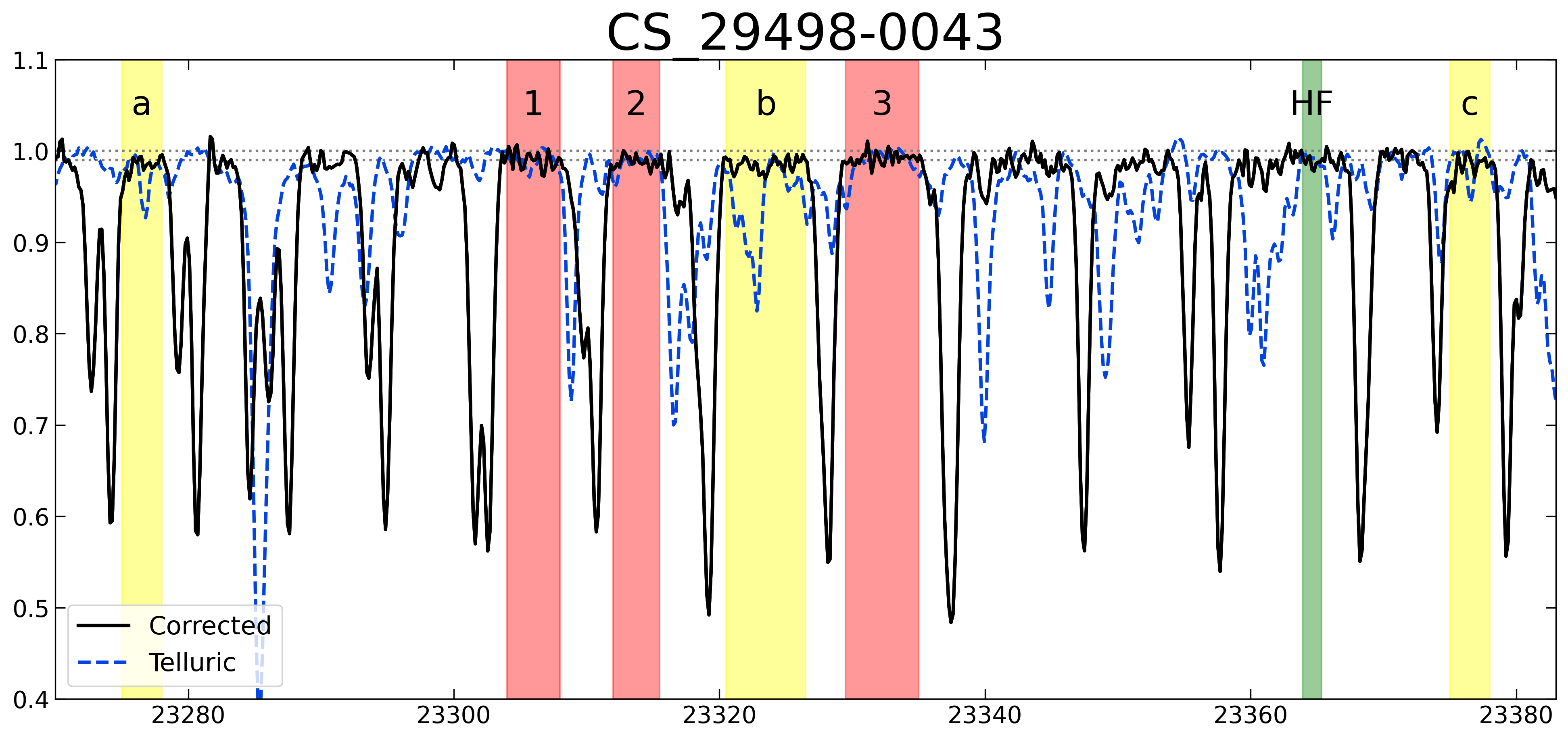}
                \includegraphics[width=.9\linewidth]{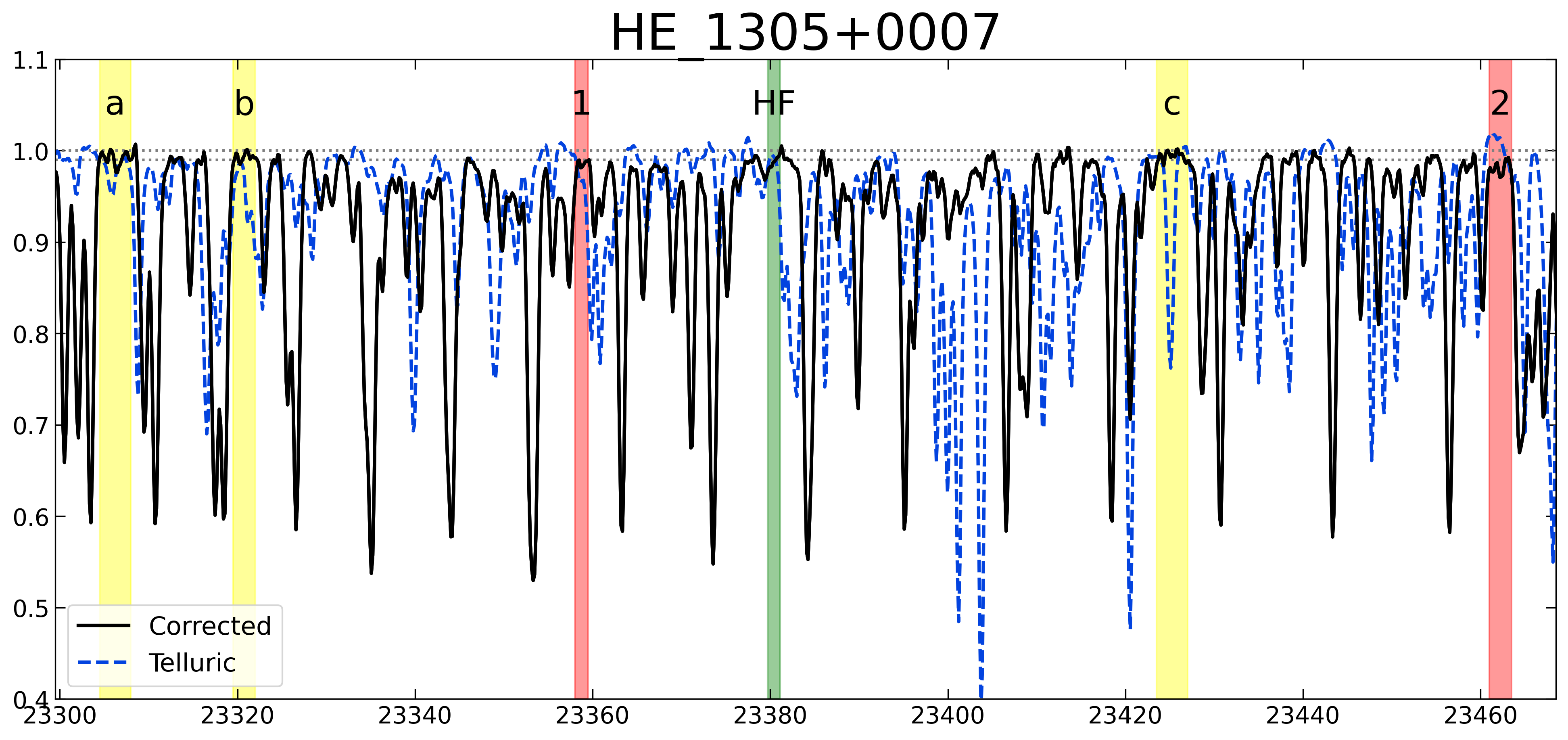}
				\caption{Observed spectra for CS 29498-0043 (upper panel) and HE 1305+0007 (lower panel). In both panels, the different lines represent the telluric corrected spectra and the telluric standard star in solid black lines and dashed blue lines, respectively. The highlighted regions indicate the recovered continuum regions after the telluric correction in yellow (a, b, c), and the telluric-free continuum regions in red (1, 2, 3). The HF absorption lines are indicated within the green marked region -- See also Table \ref{t:contin}.} \label{fig:rtell}
\end{center}
\end{figure*}

\begin{table*}
    \begin{center}
        \begin{threeparttable}
\caption{Results from the analysis of F abundances in the observed sample, including detections and upper limits. Information of the object in columns 1-2; absolute and relative abundances in  columns 3-4; predicted \FFe\ and model prediction reference in columns 5-6.}\label{t:Fabundances}
\begin{tabular}{
l
c
r@{\hspace{1mm}}l
r@{\hspace{1mm}}l
c
c}
\toprule
Star ID       & Type &  \multicolumn{2}{c}{A(F)} &  \multicolumn{2}{c}{[F/Fe]} & [F/Fe]$_{\rm Model}$ & Model ref.$^*$	\\
\midrule
CS 29498-0043 & no   &  +2.69 & $\pm$ 0.4 &  +2.00 & $\pm$ 0.4 & $-0.25$ & \citetalias[][Model A]{Tominaga2014}   \\
              &      &        &           &        &           & $+3.02$ & \citetalias[][Model B]{Tominaga2014} \\
              &      &        &           &        &           & $+2.28$ & Model B2 \\
              
HE 1305+0007  & r/s  &  +3.18 & $\pm$ 0.3 &  +0.90 & $\pm$ 0.3 & $+2.28$ & \citetalias{Abate2015b}      \\
HE 1523-1155  & s    & <+3.46 &           & <+1.10 &           & $+2.24$ & \citetalias{Abate2015b}      \\
CS 29502-0092 & no   & <+2.76 &           & <+1.50 &           &  ---  &  ---   \\
HD 126587     & no   & <+2.47 &           & <+1.00 &           &  ---  &  ---   \\
HE 0414-0343  & r/s  & <+3.00 &           & <+1.30 &           &  ---  &  ---   \\
HE 1116-0634  & no   & <+1.83 &           & <+1.00 &           &  ---  &  ---   \\
\hline
\end{tabular}

        \begin{tablenotes}
        \item[$^*$] T14= \cite{Tominaga2014}, models A and B; A15b=\cite{Abate2015b}, F abundance by private communication.
        \end{tablenotes}

        \end{threeparttable}
    \end{center}

\end{table*}

\begin{table*}
    \begin{center}
        \begin{threeparttable}
\caption{Comparison of the central depth of the HF molecular line and standard deviation in continuum regions of the spectra. The selected regions correspond to recovered continua from telluric-corrected regions, and continua from telluric-free regions presented with letters and numbers, respectively, as indicated in Fig. \ref{fig:rtell}.}\label{t:contin}
\begin{tabular}{lccccccccc}
\toprule
        &  HF  &  \multicolumn{4}{c}{$\sigma$}  &  \multicolumn{4}{c}{$\sigma$} \\
Object  &  central  &  \multicolumn{4}{c}{Telluric Regions}  &  \multicolumn{4}{c}{Telluric-free Regions} \\
   \cmidrule(l){3-6}    \cmidrule(l){7-10}
              &  depth  & a      &   b    &   c    & avg.  &   1    &   2    &   3   & avg.  \\
\midrule
CS 29498-0043 &  0.019  & 0.010  & 0.007  & 0.011  & 0.009 & 0.009  & 0.006  & 0.007 & 0.007 \\
HE 1305+0007  &  0.012  & 0.007  & 0.005  & 0.005  & 0.006 & 0.008  & 0.007  & ---   & 0.007 \\
\bottomrule
\end{tabular}
        \end{threeparttable}
    \end{center}
\end{table*}

Fig. \ref{fig:rtell} displays the corrected (solid black lines) and uncorrected (dashed blue lines) spectra for telluric absorption in CS 29498-0043 (upper panel) and HE 1305+0007 (lower panel).
To quantify the significance of our detections in these two stars, two sets of continuum regions were compared to the HF absorption. The first set of continua are those recovered after the telluric removal. The second set corresponds to those regions free of telluric absorption. 
Fig. \ref{fig:rtell} displays the selected set of continua in CS 29498-0043 and HE 1305+0007 highlighted in yellow (a, b, c) and red (1, 2, 3) for the recovered and telluric-free continuum regions, respectively.
The average standard deviation in both sets is then compared to the central depth of the HF line to test its significance, presented in Table \ref{t:contin}. 
The rest of the sample shows strong telluric lines overlapping on top of HF and/or no evident detections.

We consider each star separately. For HE 1305+0007, the average standard deviation of the telluric-corrected ($\sigma_{\rm{tc}}$) and telluric-free ($\sigma_{\rm{tf}}$) regions is very similar, $\sigma_{\rm{tc}}=0.006$ and $\sigma_{\rm{tf}} = 0.007$, respectively. The comparison of the average value of those continuum regions $\sigma_{\rm{cont}}=0.0065$, to the central depth of the HF line, HF$_{\rm{depth}}=0.012$, is used here as an argument for the significance of the detection; i.e. $0.012 / 0.0065 \simeq~2\sigma$. Similarly, for CS 29498-0043, the average standard deviations for the set of continua in the telluric-corrected and telluric-free regions are $\sigma_{\rm{tc}} = 0.009$ and $\sigma_{\rm{tf}} = 0.007$, respectively. The average of these values, $\sigma_{\rm{cont}}=0.008$, compared with the central depth of the HF line, HF$_{\rm{depth}}=0.019$, in CS 29498-0043, indicate a significance in the detection of  $0.019 / 0.008 \simeq~2\sigma$. While ideally even higher SNR spectra would be obtained for these objects, in the following sections the 2-sigma HF detections are assumed to be present in both objects.

%%%%%%%%%%%%%%%%%%%%%%%%%%%%%%%%%%%%%%%%%%%%%
\section{Results}\label{Results_CEMP2}
%%%%%%%%%%%%%%%%%%%%%%%%%%%%%%%%%%%%%%%%%%%%%

The results from the analysis of F abundances observed in our sample of 7 CEMP stars are reported in Table \ref{t:Fabundances}. In that table, theoretical predictions for F abundances are included for comparison, when available. The predicted F abundances are adopted from \cite{Tominaga2014} for Pop III SN nucleosynthesis, and from \cite{Abate2015b} for AGB nucleosynthesis and binary evolution. In the sample, the results for HE 1305+0007 are obtained from additional observations in February 2020 (see Table \ref{t:obs_CEMP2} in Sect. \ref{Observations_CEMP2}). CS 29502-0092 has been previously observed by \cite{Lucatello2011} from which we compare our results.\\

The abundance analysis indicates 2-sigma detection of F abundances in stars CS 29498-0043, \FFe$_{2\sigma} = +2.0\pm0.4$, and HE 1305+0007, \FFe$_{2\sigma} = +0.9\pm0.3$. The consideration of velocities from Earth's orbital motion in the design of the new observations for these two stars provided good results eluding telluric lines as presented in Fig. \ref{fig:detection}. For HE 1305+0007, this updated result is in excellent agreement with what was found in \citetalias{Mura2020}. For the rest of the sample, only upper limits are reported due to the absence of the absorption feature.  Abundance errors due to uncertainties in stellar parameters are shown in Table \ref{t:errors_cmp2}.

\begin{figure*}
\begin{center}
    \includegraphics[width=\linewidth]{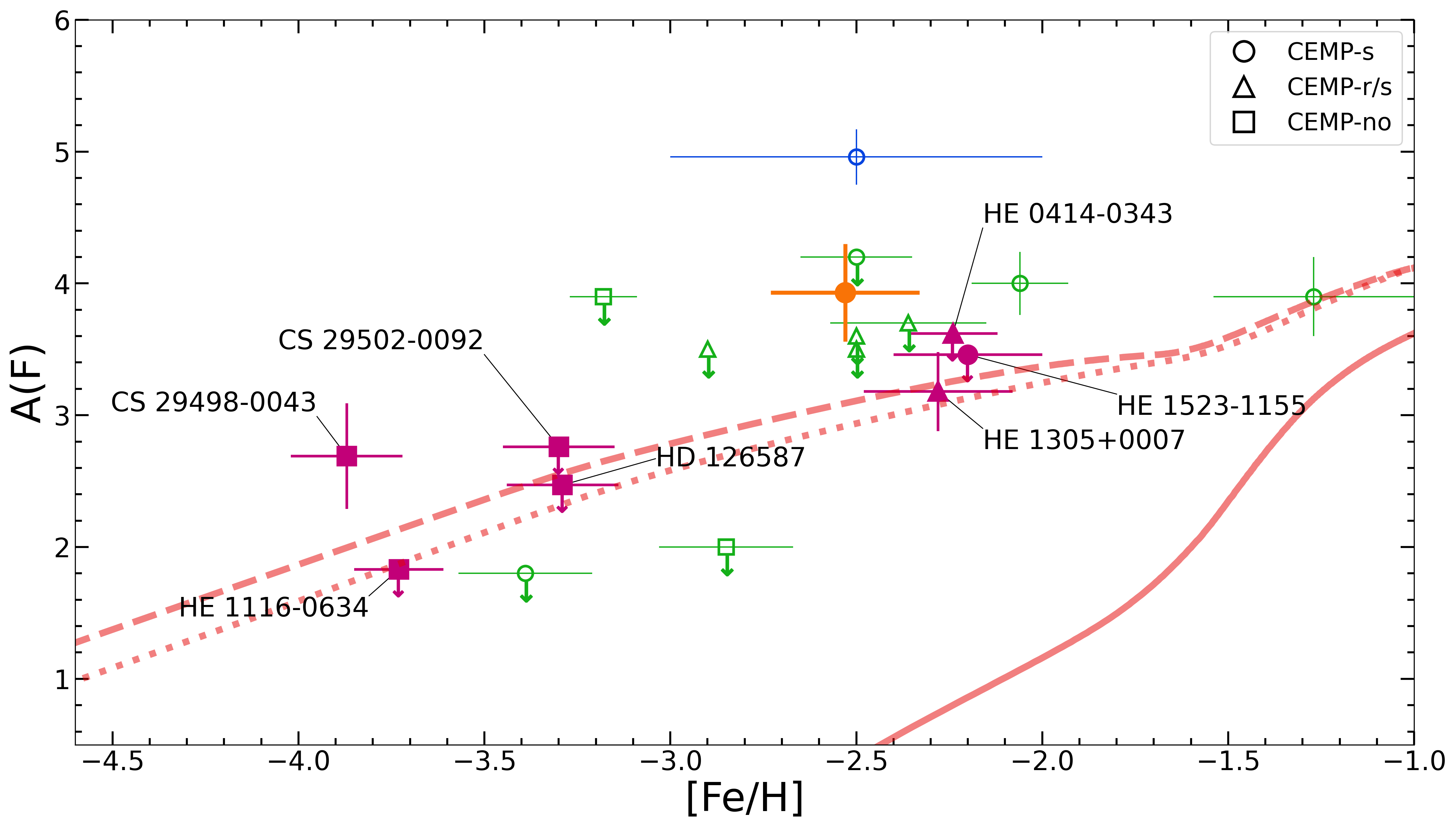}
    
    \caption{Current state of the observational data for A(F) vs. [Fe/H] at low metallicities. Literature data: \citet{Lucatello2011}, in green; \citet{Schuler2007}, in blue. F measurement in HE 1429-0551 from \citetalias{Mura2020}, is shown in orange. Abundances and upper limits obtained in this work are presented in magenta. The red lines show Galactic chemical evolution models for the solar neighborhood from \citet[][solid line]{Kobayashi2020}, with Wolf--Rayet winds (dotted line), and fully with rotating massive stars (dashed line) from \citet{Limongi2018}.}
    \label{fig:F_literature}
\end{center}
\end{figure*}

Fig. \ref{fig:F_literature} is our updated version of fig. 3 presented in \citetalias{Mura2020}. The measurements and upper limits obtained from the new observations are presented in magenta: CEMP-$s$ as circles, CEMP-$r/s$ as triangles, and CEMP-$no$ as squares. The metallicity errors displayed in the figure correspond to those reported from the references for each star (see Table \ref{t:stellar_param_cemp2}).

\begin{table}
    \begin{center}
        \begin{threeparttable}
\caption{Fluorine abundance dependence on stellar parameter uncertainties for CS 29498-0043 and HE 1305+0007.}\label{t:errors_cmp2}
\begin{tabular}{l c@{\hspace{2mm}} c@{\hspace{2mm}} c@{\hspace{2mm}} c@{\hspace{2mm}} c}
\toprule
Star ID         &   $\Delta$\Teff (K)&  $\Delta$\logg&  $\Delta$\Fe&    $\Delta$\micro (\velo)& Total$^{\star}$\\
                &   +100             &  +0.50        &  +0.30      &    +0.30                 &                \\
\midrule
CS 29498-0043   &   +0.30            &  -0.10        &  -0.30      &    +0.00                 & 0.44           \\
HE 1305+0007    &   +0.30            &  +0.05        &  -0.10      &    +0.05                 & 0.32          \\

\bottomrule
\end{tabular}

        \begin{tablenotes}
        \item[$\star$] The total value is the quadrature sum of the individual abundance dependencies.
        \end{tablenotes}

        \end{threeparttable}
    \end{center}
\end{table}

%#######################
% Fig F at low metallicities
%#######################

%%%%%%%%%%%%%%%%%%%%%%%%%%%%%%%%%%%%%%%%%%%%%
\section{Discussion}\label{Discussion_CEMP2}
%%%%%%%%%%%%%%%%%%%%%%%%%%%%%%%%%%%%%%%%%%%%%

These new F abundances, in addition to the larger number of upper limits, expand the observational framework of CEMP stars in the number of observed objects, types, and metallicity range. Given the scarce availability of observed F abundances at low metallicities, a larger sample will benefit the theoretical community by imposing new constraints to nucleosynthesis in stars and supernovae. The production of F via AGB nucleosynthesis has been intensively discussed in \citetalias{Mura2020}. Here we discuss our observational results with IGRINS at Gemini-S, as well as with the literature data, comparing to the theoretical works from \cite{Abate2015b} and \cite{Tominaga2014}.

%%%%%%%%%%%%%%%%%%%%%%%%%%%%%%%%%%%%%%%%%%%%%
\subsection{F in CEMP-$s$ and CEMP-$r/s$ stars}

As discussed in \cite{Frebel2015}, \textit{stellar archaeology} is the use of stellar chemical abundances in the most metal-poor stars to study the astrophysical sites, conditions of nucleosynthesis, and the major physical processes that drove early star formation. In this context, AGB stars at low metallicity have short lifetimes and cannot be directly observed. However, nucleosynthesis in AGB stars at low metallicity can be inspected through CEMP stars formed via mass transfer scenario. The theoretical models from \cite{Abate2015b} attempt to reproduce the observed abundances in CEMP-$s$ and CEMP-$r/s$ stars in this context. Although their models can reproduce observed abundances in CEMP-$s$ stars, they over-predict light element abundances for CEMP-$r/s$ stars. As shown in \citetalias{Mura2020}, F abundances in these objects can be used as a test and provide additional constraints to the AGB nucleosynthesis. From the observed sample in this work, stars HE 1305+0007 and HE 1523-1155 are included in \cite{Abate2015b}, and therefore F predictions are available for AGB nucleosynthesis at low metallicities and binary evolution.\\

HE 1305+0007, a CEMP-$r/s$ star, was included in the analysis of \citetalias{Mura2020}. The observations of HE 1305+0007 obtained in 2019 were not of sufficiently high quality (e.g., SNR, severe telluric contamination) in the region of HF to determine its F abundance, instead, an upper limit of $\FFe < +1.0$ was suggested \citepalias[see fig. 2,][]{Mura2020}. The new observations of HE 1305+0007 from IGRINS at the Gemini-S telescope, in conjunction with a careful observational time constraint to avoid telluric contamination on top of the HF line, delivered high-quality data from which F was detected, $\FFe=+0.9$. This new result is in excellent agreement with the previous upper limit and reaffirms the discrepancies between observations and the predicted F abundance, $\FFe_{Abate}=+2.27$. The conflicting abundances are not unique to F. In order to reproduce the enhanced abundances in the $s$- and $r$-process elements simultaneously, the model tends to overestimate the abundances of light elements, such as C, N, F, Na, and Mg \citep{Abate2015b}. Although the origin of CEMP-$r/s$ stars and their nucleosynthesis are still unclear, these new results on F abundances in HE 1305+0007 will help to elucidate between different proposed progenitors. For particular sets of conditions, the intermediate neutron-capture ($i$-) process with densities of the order $n$ $\simeq$ $10^{15}$ cm$^{-3}$ \citep{Cowan1977} offers an alternative description of the chemical profile in HE 1305+0007. \cite{Hampel2016,Hampel2019} well-reproduced the observed abundances in this star, including C, $s$- and $r$-process elements \citep[see also][for additional recent predictions and yields from $i$-process]{Choplin21}. The F abundances presented here will help to test and constrain new formation and nucleosynthesis channels, such as the $i$-process.\\

The upper limit derived from the observations in the CEMP-$s$ star HE 1523-1155 suggests a significantly lower F content, $\FFe<+1.10$, compared to the expected value from \cite{Abate2015b}, $\FFe_{Abate}=+2.24$. The disagreement in the predicted F abundance may be due to multiple sources of uncertainties in the model, from AGB nucleosynthesis (e.g., uncertain nuclear reaction rates) to the binary evolution (e.g., unknown orbital period).\\

The CEMP-$r/s$ star HE 0414-0343 is not included in \cite{Abate2015b}. However, \cite{Hollek2015} suggested a late-time mass transfer from a 1.3 \Msun\ AGB star for the observed chemical pattern in HE 0414-0343. Their models can reproduce the abundances in C and N, as well as the $s$- and $r$-process elements within an order of magnitude. The derived F abundance from the model indicates $\rm{A(F)} \sim +2.0$, lower than the upper limit presented in this work, $\rm{A(F)} < +3.0$. Therefore, this result cannot provide strong constraints to the models from \cite{Hollek2015} on HE 0414-0343.\\

Additional observations on a set of well-chosen targets, for which F measurements are most likely (see sect. \ref{detection_limit}), will help to strengthen the observational sample. In turn, a larger observational sample of F abundances at low metallicities may motivate additional theoretical efforts for F production from AGB stars in the early Galaxy.

%%%%%%%%%%%%%%%%%%%%%%%%%%%%%%%%%%%%%%%%%%%%%
\subsection{F in CEMP-$no$ stars}

%#######################
% Fig F at low metallicities
%#######################
\begin{figure}
\begin{center}
    		\includegraphics[width=\linewidth]{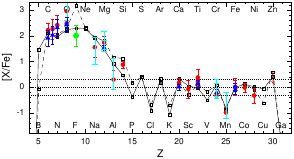}
				\caption{Updated abundance profile for CS29408-0043 with our F measurement (green). Open black squares with a dashed line represent the previous Model B and the open black squares with a solid line represent the updated version, Model B2, with a lower explosion energy, E$_{51} = 10$. Filled red circles and blue triangles are observational data as in {\citet[][see the references therein]{Tominaga2014}}.
                }
    \label{fig:new_CS29498}
\end{center}
\end{figure}

The formation mechanisms for CEMP-$no$ stars are still unclear. From the proposed scenarios that explain the C enhancements in CEMP stars, faint SN can reproduce the chemical enrichment observed in CEMP-$no$ stars through only one Pop III SN event. Alternative models can also reproduce the CEMP-$no$ chemical pattern but require multiple events occurring in a short period of time to fit the observed CNO and Fe-peak abundances \citep[e.g.,][]{Meynet2006,Limongi2018}. 
\cite{Hartwig2023} used machine learning techniques to explore the fraction of extremely metal-poor stars born in a mono- or multi-enriched environment. The data-driven method included 462 analyzed stars from several spectroscopic data including the SAGA database and 30 additional stars from \cite{Ishigaki2018} as well as the most metal-poor stars to date \citep[see section 2.1 in][]{Hartwig2023}. In their work, this method suggests that $\sim 30\%$ of the sample is mono-enriched at $\Fe = -3.50$. This fraction of mono-enrichment, strongly increases at $\Fe < -4.0$, and most of the mono-enriched stars are CEMP stars (see their fig. 6). The probability of mono-enrichment of CS 29498-0043 is $92\pm5$\% (table 2 from \citealt{Hartwig2023}).

\cite{Tominaga2014} used models of faint SN from Pop III stars to conduct \textit{abundance profiling} of a sample of CEMP-$no$ stars, including CS 29498-0043. In their work, the chemical abundances in CS 29498-0043 are reproduced by two models, A and B, previously described in Section \ref{Target_Selection_Overview} \citep[see also][section 2]{Tominaga2014}. The main difference in both models is the different mixing efficiency during hydrostatic burning as prescribed by \cite{iwamoto2005} -- i.e., Model A without enhanced mixing; and Model B with enhanced mixing.

In massive stars, the mixing efficiency is suggested to be modulated by rotation, with larger values of rotation producing enhancements in the surface abundances from material processed by hydrogen burning, such as N \citep{Heger2000, Meynet2000,Chieffi2013}. In the context of the most metal-poor stars, observations of massive stars support the primary production of N, that is, the [N/H] ratios scale roughly with [Fe/H]. One way to explain such observations is through fast-rotating massive stars, or spinstars, \citep{Hirschi2007,Maeder2015}. 

The enhanced N can be converted to F if the N-rich layer experiences high temperature ($>7\times10^8$ K) during the shock propagation (Shibata et al. in prep.).
While \citet{Limongi2018} found that the majority of F is produced in He-shell burning by $^{14}$N($\alpha$, $\gamma$)$^{18}$F($\beta^+$)$^{18}$O(p, $\alpha$)$^{15}$N($\alpha$, $\gamma$)$^{19}$F,
in \cite{Tominaga2014}, the majority of F is formed during explosive burning by the hot CNO reaction \chem{14}{N}($p,\gamma$)\chem{15}{O}($\alpha,\gamma$)\chem{19}{Ne}($\beta +$)\chem{19}{F}, when the shock propagates through the N-rich H layer.
Since the enhancement in F abundance is dependent on the temperature achieved in the N-rich layer during the shock wave; we can find this only for hypernovae with N-rich H layers due to the enhanced mixing. Therefore F abundances can provide insights to constrain the explosion energies in SN and other parameters, such as rotation.
The models used to describe the abundance profile of CS 29498-0043 have the same energy explosion, E$_{51}=20$, because of high [Co/Fe] and the difference between the models lies primarily in the enhanced mixing from which the predicted F abundances are strikingly different, $\FFe_{Model\,A} = -0.25$ and $\FFe_{Model\,B} = +3.02$ \citep{Tominaga2014}. Our measured F abundance in CS 29498-0043 is $\FFe = +2.0 \pm 0.4$. This result is significantly higher than in Model A (without enhanced mixing) and supports the prediction from Model B (with enhanced mixing), although F is overproduced by $\sim 1$ dex. One possibility would be that the observed F abundance rejects the model from \cite{Tominaga2014} with no enhanced mixing (Model A), and that the abundance profile requires a \textit{moderate} amount of mixing, at a level below their Model B, with enhanced mixing or a lower explosion energy. 

Fig. \ref{fig:new_CS29498} presents an updated abundance profile of CS 29498-00043 which includes our new value on F abundance ($\FFe =+2.0$) in this star. The new Model B2 considers another possibility, i.e., a lower explosion energy of a hypernova, assuming (I) explosion energy E$_{51}=10$, (ii) final central remnant mass M$_{\rm rem}=4.95$ \Msun, (iii) ejected Fe mass M$_{\rm ej}({\rm Fe})=1.19 \times 10^{-3}$ \Msun, and (iv) “low-density” factor $f_\rho=1/6$. The main differences between the previous and the updated chemical profile for CS29498-0043 are visible especially in F, which is now $\FFe_{Model\,B2} = +2.28$, where most of F is synthesized by the explosive nucleosynthesis. While the amount of F in the progenitor is only $5.74\times10^{-8}$ \Msun, the ejected amount of F is $8.88\times 10^{-5}$ \Msun. This demonstrates that the F abundance is a sensitive indicator of explosion energy if the N is already enhanced in the progenitor star. 

The characterization of faint supernova/hypernova explosion modeling is a complex task. Temperature, entropy, and fallback are determined by energy injection, geometry and the fraction of kinetic energy in the jet-induced explosion \citep{Tominaga2014,Tominaga2009}. However, the arbitrary choice of mixing-and-fallback parameters and the low-density modifications in the models led to weak constraints on energy explosion. The observation of F abundance in CS 29498-0043 serves as a direct test to models and offers valuable new constraints (e.g., energy explosion) for our understanding of the first chemical enrichment mechanisms.\\

%%%%%%%%%%%%%%%%%%%%%%%%%%%%%%%%%%%%%%%%%%%%%
\subsection{F production from Wolf--Rayet stars in Galactic chemical evolution}

In addition to CS 29498-0043, upper limits were derived in three more CEMP-no stars, i.e., CS 29502-0092, HD 126587, and HE 1116-0634 which were not examined in \cite{Tominaga2014}. \cite{Lucatello2011} also obtained an upper limit of F abundance in CS 29502-0092 that can be compared to our results. In their analysis, \cite{Lucatello2011} use the stellar parameters: \Teff\ $=4890$, \logg\ $=1.7$, and \Fe\ $=-3.18$, reported to be adopted from \cite{Lai2007}. However, \cite{Lai2007} presents different values for that star: \Teff\ $=5114$, \logg\ $=2.51$, and \Fe\ $=-2.92$. The derived upper limit by \cite{Lucatello2011} for CS 29052-0092 is $\FFe <+2.5$ (A(F) $ < +3.5$), $\sim1.5$ dex higher than our A(F) results presented here.\\

At $\Fe < -3.5$, the probability of mono-enrichment increases even for CEMP stars, and the abundance of stars might reflect the chemical evolution of the Galaxy. Since CEMP-$s$ stars are further enriched by mass-transfer from an AGB companion, their original chemical composition is altered by the accreted material, no longer reflecting the composition from their formation gas cloud. Therefore, we compare only CEMP-no results to Galactic chemical evolution models of F towards the extremely low-metallicity regime.
In Fig. \ref{fig:F_literature} red lines show Galactic chemical evolution models of the solar neighborhood, as presented by \citet[][solid line]{Kobayashi2020}, along with updated models incorporating Wolf–Rayet winds (dotted line) or fully integrating rotating massive star yields from \citet[][dashed line]{Limongi2018}.
While massive stars yields from \citet{Limongi2018} moderately overproduce F for HE 1116-064 and marginally for HD 126582, both, Wolf-Rayet and rotating massive stars models are in reasonable agreement with our derived upper limits. However, While upper limits are useful to describe the maximum possible abundance for a certain star, in order to constrain the contribution and distribution of stellar rotational velocities, it is necessary to increase the sample of measurements CEMP-no stars.

%%%%%%%%%%%%%%%%%%%%%%%%%%%%%%%%%%%%%%%%%%%%%
\subsection{Detection Limit for F Abundances at Low Metallicities}\label{detection_limit}

It is clear that F abundances can provide strong constraints and a deeper insight into the nucleosynthesis processes and mechanisms at the low metallicity regime. However, the low number of detections and upper limits at low metallicities remains small and more data are needed to provide a solid observational framework to compare with theoretical models. This observational deficit is due to the difficulties in the detection of F abundances via HF from the 23358.329 \AA\ line, which becomes even harder at low metallicities, $\Fe<-2.0$. In addition, the strength of HF (1-0) R 9, at 23358.329 \AA, is highly sensitive to \Teff, and also \logg\ to a lesser extent. At a fixed metallicity, lower temperature and lower surface gravities increase the strength of the HF molecular lines. When the temperature is lower, the density of molecules is higher, while lower surface gravity amplifies the strength of the lines compared to the continuum \citep[see e.g.,][for a discussion on the matter]{Jonsson2014a,Ryde2020}. In addition, and as previously discussed, small remnants from the telluric correction process have the potential to distort or eliminate any signal from a weak HF absorption, such as is the case for metal-poor stars.

Prior to this work, four F detections have been reported in CEMP stars \citep{Schuler2007,Lucatello2011,Mura2020}. Our results increased the number of detections by 66\%. Additionally, our lowest F abundance measured in CS 29408-0043 at [Fe/H] = $-3.87$, is more than a factor of 10 lower in metallicity than the closest detection \citepalias[HE 1429-0551 at $\Fe = -2.53$,][]{Mura2020}.\\

\begin{figure}
\begin{center}
    		\includegraphics[width=\linewidth]{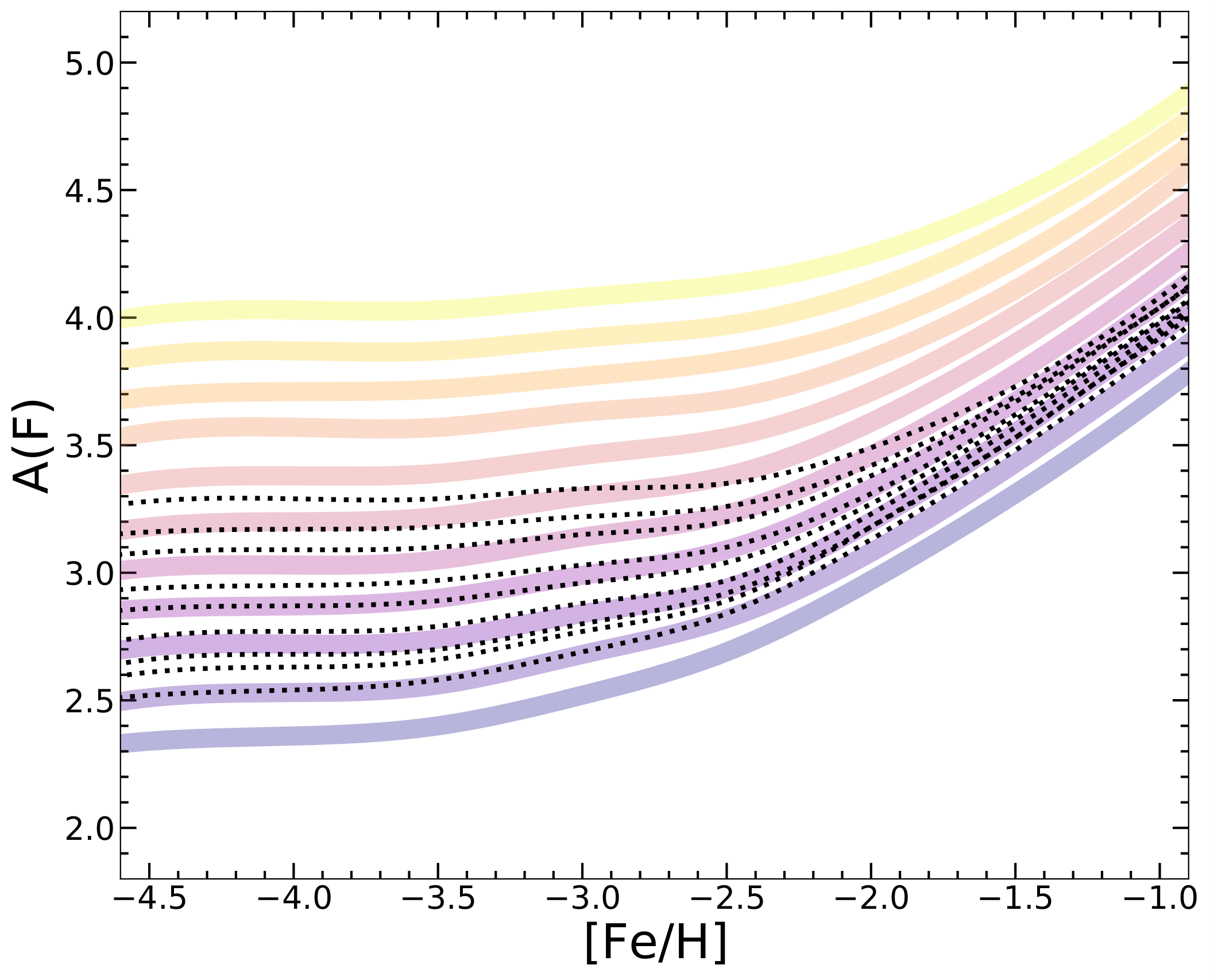}
				\caption{Approximate detection thresholds for F abundances, as a function of metallicity, to reach 2\% molecular absorption relative to the continuum from the HF (1-0) R 9 line, at 23358.329 \AA. The F abundances were derived using synthetic spectra and stellar atmosphere models at different effective temperature and surface gravity values. Color-scaled lines present the F abundances obtained using \Teff\ from 4400K to 4900K (bottom purple line and top yellow line, respectively) with fixed \logg\ $= 1.2$. Each line corresponds to a 50K difference from its neighbor line. Similarly, black dotted lines show the F abundances for a range of surface gravities from \logg\ $= 0.1$, at the bottom, to \logg\ $= 1.7$ at the top, with fixed \Teff\ $= 4500$. The \logg\ difference between each neighbor line is 0.2 dex.} \label{fig:F_Threshold}
\end{center}
\end{figure}

Detection thresholds for F abundance determination provide extremely valuable information for selecting future targets for study. Fig. \ref{fig:F_Threshold} displays the detection thresholds for F abundances to reach HF absorption of approximately 2\% relative to the continuum using synthetic spectra and stellar atmosphere models within a range of \Teff\ and \logg. In that figure, the variations in \Teff\ are presented using color-scaled lines from 4400K, in purple at the bottom,  up to 4900, in yellow at the top. Each line corresponds to a 50K increment in \Teff\ at a fixed surface gravity of \logg\ = 1.2. The black dotted lines represent the variations in surface gravities, at a fixed \Teff\ = 4500, starting at \logg\ = 0.1 at the bottom, up to \logg\ = 1.7 at the top, in steps of 0.2 dex for each line. At low metallicities, the F abundance needed to reach the detection limit increases significantly for higher temperatures. The differences in abundance become smaller towards higher metallicities. These abundance differences are also observed for the variations in surface gravities, although, to a lower extent.\\

Different components play a role in the strength of the HF absorption such as the line and continuum opacity ratio, electron pressure, and molecular equilibria. The behavior of HF and other hydride molecules at low metallicities has been addressed and described by \cite{Cottrell1978}.

%#######################
% Fig F at low metallicities
%#######################

%%%%%%%%%%%%%%%%%%%%%%%%%%%%%%%%%%%%%%%%%%%%%
\section{Conclusions}\label{Conclusions_CEMP2}
%%%%%%%%%%%%%%%%%%%%%%%%%%%%%%%%%%%%%%%%%%%%%

This work has provided a pioneering and pilot study of the use of fluorine abundances in extremely metal-poor stars as a tool to probe the nature of the first stars. In particular, the CEMP-no class of objects continues to pose questions regarding their formation mechanisms. The abundance profiling study by \cite{Tominaga2014}, provided distinct predictions for the [F/Fe] ratio for one CEMP-no object, CS 29498-0043, for which a F measurement of $\FFe = +2.0$ (A(F) = $+2.69$) was obtained in this study. Our F measurement at $\Fe = -3.87$ is the F measurement at the lowest metallicity obtained to date. The comparison of the F measurement with the predictions from \cite{Tominaga2014} supports the progenitor model in which there is enhanced mixing in pre-supernova stars, which could also be due to stellar rotation.
Since our measured value remains significantly below the \citeauthor{Tominaga2014}'s original prediction with enhanced mixing ([F/Fe] = $+3.02$), %this result strongly supports the need for 
a moderate amount of mixing in the progenitor star, and/or a lower explosion energy of hypernovae, is required. In fact, the updated version of the theoretical model for CS 29498-0043 from \cite{Tominaga2014} with a lower energy explosion (E$_{51}=10$) can also describe our observed F value without departing significantly from the rest of the observed abundance profile.

It is important to note that in this model, the majority of F is produced in the N-rich H-layer when the supernova shock propagates.
This is different from the F production in He shell burning in Wolf-Rayet stars \cite[e.g.,][]{Limongi2018}. During supernova explosion, further enhancement by neutrino process is also expected \cite[e.g.,][]{Kobayashi2011b}.
In order to determine which is the major production site, it is necessary to use a multi-dimensional explosion model of a rotating star with neutrino process and updated nuclear reaction rates.\\

The detection of F abundance A(F) = $+3.18$ ($\Fe = +0.90$) in the CEMP-$r/s$ stars, HE 1305+0007, is in excellent agreement with the upper limit found in \citetalias{Mura2020} A(F) < $+3.28$ ($\Fe < +1.00$). Additional upper limits were derived for the rest of the sample of CEMP-$s$ and CEMP-$r/s$. In general, the theoretical models from \cite{Abate2015b} over-predict the F abundances in these objects. The over-predicted F abundances may indicate a less efficient production of $^{19}$F in their progenitors but also restate the possibility for alternative progenitors and nucleosynthesis processes such as the $i$-process \citep{Hampel2016,Hampel2019,Choplin21}.\\

The contribution from Wolf-Rayet stars in the chemical enrichment history of the Galaxy is also discussed in this paper.
A moderate rotation/mixing seems also to be supported, but a larger sample of measurements is required to draw a conclusion. This work has increased the number of detections of F abundance at metallicities $\Fe <-2.0$ by 66\%. Additionally, the entire observed sample in this work (including detections and upper limits), has increased the current literature data by 63\%.\\

The two measurements obtained in this work were possible due to the design of the observations for these two objects. The scheduled observations in which Earth's orbital motions were considered to shift the HF region into telluric-free regions, gave excellent results. Future observations to obtain F abundance at low metallicities would benefit should consider this factor, when possible, and consequently, strengthen the observational framework of F abundances at these metallicities.

%%%%%%%%%%%%%%%%%%%%%%%%%%%%%%%%%%%%%%%%%%%%%
\section*{Acknowledgements}
%%%%%%%%%%%%%%%%%%%%%%%%%%%%%%%%%%%%%%%%%%%%%

The authors sincerely thank the reviewer for the constructive comments and suggestions that helped clarify this article.
Most of this work was done on the traditional lands of the Ngunnawal and Ngambri people of the Canberra region. AMG acknowledges and pays respects to their elders, present and emerging.
AMG appreciates the fruitful discussions with Thomas Nordlander, Paula Jofré, Melanie Hampel, and Simon Campbell for providing feedback and suggestions on the draft of this paper. AMG also thanks the partial financial support of FONDECYT Regular grant number 1200703 and the support by ANID (Former CONICYT; Chile) through Programa Nacional de Becas de Doctorado 2017 (CONICYT-PCHA/Doctorado Nacional/2017- 72180413).
NT appreciates Masaki Shibata for the investigation of nuclear processes to synthesize F during the explosive nucleosynthesis.
CK acknowledges funding from the UK Science and Technology Facility Council through grants ST/R000905/1, ST/V000632/1, ST/Y001443/1, and also the Stromlo Distinguished Visitorship at the ANU.
This work was supported by the Australian Research Council Centre of Excellence for All Sky Astrophysics in 3 Dimensions (ASTRO 3D), through project number CE170100013.
%
%telescopes/institutions
This work used the Immersion Grating Infrared Spectrometer (IGRINS) that was developed under a collaboration between the University of Texas at Austin and the Korea Astronomy and Space Science Institute (KASI) with the financial support of the US National Science Foundation 27 under grants AST-1229522 and AST-1702267, of the University of Texas at Austin, and of the Korean GMT Project of KASI.

The results of this work are based on observations obtained at:
i) the international Gemini Observatory, a program of NSF NOIRLab, which is managed by the Association of Universities for Research in Astronomy (AURA) under a cooperative agreement with the U.S. National Science Foundation on behalf of the Gemini Observatory partnership: the U.S. National Science Foundation (United States), National Research Council (Canada), Agencia Nacional de Investigaci\'{o}n y Desarrollo (Chile), Ministerio de Ciencia, Tecnolog\'{i}a e Innovaci\'{o}n (Argentina), Minist\'{e}rio da Ci\^{e}ncia, Tecnologia, Inova\c{c}\~{o}es e Comunica\c{c}\~{o}es (Brazil), and Korea Astronomy and Space Science Institute (Republic of Korea);
ii) the Lowell Discovery Telescope (LDT -- former Discovery Chanel telescope, DCT) at Lowell Observatory. Lowell is a private, non-profit institution dedicated to astrophysical research and public appreciation of astronomy and operates the LDT in partnership with Boston University, the University of Maryland, the University of Toledo, Northern Arizona University and Yale University.

%%%%%%%%%%%%%%%%%%%%%%%%%%%%%%%%%%%%%%%%%%%%%%%%%%
\section*{Data Availability}
%%%%%%%%%%%%%%%%%%%%%%%%%%%%%%%%%%%%%%%%%%%%%

Data is available on request. The data underlying this article will be shared on reasonable request to the corresponding author.



\bibliographystyle{mnras}
\bibliography{F_CEMP_PaperII} % if your bibtex file is called example.bib

%\begin{thebibliography}{99}
%\end{thebibliography}

%%%%%%%%%%%%%%%%%%%%%%%%%%%%%%%%%%%%%%%%%%%%%%%%%%

%%%%%%%%%%%%%%%%% APPENDICES %%%%%%%%%%%%%%%%%%%%%

%\appendix


%If you want to present additional material which would interrupt the flow of the main paper, it can be placed in an Appendix which appears after the list of references.

%%%%%%%%%%%%%%%%%%%%%%%%%%%%%%%%%%%%%%%%%%%%%%%%%%

% Don't change these lines
\bsp	% typesetting comment
\label{lastpage}
\end{document}